\documentclass[aps,twocolumn,superscriptaddress,floatfix]{revtex4}
\usepackage[dvips]{graphicx}
\usepackage{latexsym}
\usepackage{amsmath}
\usepackage{amsfonts}
\usepackage{amssymb}
\usepackage{bm}
\usepackage{txfonts}

\begin{document}
\newcommand{\fig}[2]{\includegraphics[width=#1]{#2}}
\def\bk{{\textbf{k}}}
\def\bq{{\textbf{q}}}
\def\bQ{{\textbf{Q}}}
\def\bM{{\textbf{M}}}
\def\bG{{\textbf{G}}}
\def\br{{\textbf{r}}}
\def\ba{{{\bm a}}}
\newcommand{\llangle}{{\langle\!\langle}}
\newcommand{\rrangle}{{\rangle\!\rangle}}
\newcommand{\avs}{$A$V$_3$Sb$_5$}
\newcommand{\kvs}{KV$_3$Sb$_5$}
\newcommand{\cvs}{CsV$_3$Sb$_5$}

\title{Loop-current charge density wave driven by long-range Coulomb repulsion on the kagom\'e lattice}

\author{Jin-Wei Dong}
\affiliation{CAS Key Laboratory of Theoretical Physics, Institute of Theoretical Physics, Chinese Academy of Sciences, Beijing 100190, China}
\affiliation{School of Physical Sciences, University of Chinese Academy of Sciences, Beijing 100049, China}

\author{Ziqiang Wang}
\thanks{Corresponding author: wangzi@bc.edu}
\affiliation{Department of Physics, Boston College, Chestnut Hill, MA 02467, USA}

\author{Sen Zhou}
\thanks{Corresponding author: zhousen@itp.ac.cn}
\affiliation{CAS Key Laboratory of Theoretical Physics, Institute of Theoretical Physics, Chinese Academy of Sciences, Beijing 100190, China}
\affiliation{School of Physical Sciences, University of Chinese Academy of Sciences, Beijing 100049, China}
\affiliation{CAS Center for Excellence in Topological Quantum Computation, University of Chinese Academy of Sciences, Beijing 100049, China}

\date{\today}

\begin{abstract}
Recent experiments on vanadium-based nonmagnetic kagom\'e metals $A$V$_3$Sb$_5$ ($A=$ K, Rb, Cs) revealed evidence for possible spontaneous time-reversal symmetry (TRS) breaking in the charge density wave (CDW) ordered state.
The long-sought-after quantum order of loop currents has been suggested as a candidate for the TRS breaking state.
However, a microscopic model for the emergence of the loop-current CDW due to electronic correlations is still lacking.
Here, we calculate the susceptibility of the real and imaginary bond orders on the kagom\'e lattice near van Hove filling, and reveal the importance of next-nearest-neighbor Coulomb repulsion $V_2$ in triggering the instability toward imaginary bond ordered CDW.
The concrete effective single-orbital $t$-$V_1$-$V_2$ model on the kagom\'e lattice is then studied, where $t$ and $V_1$ are the hopping and Coulomb repulsion on the nearest-neighbor bonds.
We obtain the mean-field ground states, analyze their properties, and determine the phase diagram in the plane spanned by $V_1$ and $V_2$ at van Hove filling.
The region dominated by $V_1$ is occupied by a $2a_0 \times 2a_0$ real CDW insulator with the inverse of Star-of-David (ISD) bond configuration.
Increasing $V_2$ indeed drives a first-order transition from ISD to stabilized loop-current insulators that exhibit four possible current patterns of different topological properties, leading to orbital Chern insulators.
We then extend these results away from van Hove filling and show that electron doping helps the stabilization of loop currents, and gives rise to doped orbital Chern insulators with emergent Chern Fermi pockets carrying large Berry curvature and orbital magnetic moment.
Our findings provide a concrete model realization of the loop-current Chern metal at the mean-field level for the TRS breaking normal state of the kagom\'e superconductors.
\end{abstract}
\maketitle

\section{Introduction}
The nonmagnetic transition-metal kagom\'e compounds $A$V$_3$Sb$_5$ ($A=$ K, Rb, Cs) \cite{Ortiz-prm19} have attracted increasing attention since the discovery of the exotic charge density wave (CDW) with $2a_0\times2a_0$ in-plane periodicity \cite{Ortiz-prl20, Hasan-nm21, Zeljkovic-nat21, Uykur-prb21, Wilson-prx21, XHChen-prx21, Hasan-prb21, Zeljkovic-np22} below $T_{\rm cdw}$ (78$\sim$103 K) and superconductivity \cite{Ortiz-prl20, Ortiz-prm21, Yin-cpl21, Chen-nat21} below $T_c$ (0.9$\sim$2.5 K).
Central to these developments is a growing set of experiments pointing to possible time-reversal symmetry (TRS) breaking, despite the absence of local moments \cite{Kenney-jpcm21} and itinerant magnetism \cite{Ortiz-prm19, Ortiz-prm21}.
A giant anomalous Hall effect \cite{Yang-sa20, Yu-prb21} was observed below $T_{\rm cdw}$, as well as anomalous Nernst, thermal Hall, and thermoelectric transport properties \cite{nernest, seebeck}.
More direct evidence for spontaneous TRS breaking was detected by muon spin relaxation \cite{muSR1, muSR2}, optical Kerr rotation \cite{kerr, kerr-liangwu, kerr-yonezawa}, circular dichroism \cite{kerr-liangwu}, and  nonlinear transport \cite{transport}.
It was conjectured \cite{Hasan-nm21} that a candidate for the TRS breaking CDW is an exotic $2a_0\times2a_0$ loop-current (LC) order, analogous to the Varma loop-current \cite{Varma, Varma-prl}, the staggered flux \cite{staggeredflux, PALee-rmp2006}, and the $d$-density wave \cite{Nayak-prb2000, ddw} states proposed in the context of the high-T$_c$ cuprates.
A novel pair density wave order was discovered in the superconducting (SC) states and in the observed pseudogap states when superconductivity is either suppressed by a magnetic field  above the upper critical field or above SC transition temperature \cite{Chen-nat21}.
Remarkably, charge-4$e$ and charge-6$e$ flux quantization were observed in the magnetoresistance oscillations in mesoscopic ring structures of thin flake \cvs samples, providing evidence for the novel higher-charge superconductivity \cite{JiangWang-arXiv22}.

On the theoretical side, first-principle density functional theory (DFT) calculations revealed that the pristine \avs\ structure is unstable due softening of acoustic phonons near the M and L points of the Brillouin zone (BZ) boundary \cite{Tan-prl21}.
The stable ground state of the crystal has a breathing structural distortion of the kagom\'e lattice with $2a_0 \times 2a_0$ in-plane periodicity, corresponding to the Star-of-David (SD) or inverse SD (ISD) configuration, stacked with period $2a_0$ along the $c$-axis, in overall agreement with the findings of experiments \cite{Hasan-nm21, Zeljkovic-nat21, XHChen-prx21, Li-prx21, Wilson-prx21, Wenzel-xiv21, Ratcliff-prm21, Xie-prb22, Wu-prb22, Liu-nc22}.
However, the electron-phonon coupling cannot produce a CDW state that spontaneously breaks TRS \cite{Tan-prl21, Ferrari-prb22}.
Given the experimental evidence suggesting broken TRS, it is therefore important to study spontaneous TRS breaking CDW states in effective models that include the electron-electron interactions beyond DFT on the kagom\'e lattice.
The instability of the kagom\'e lattice electrons at the van Hove (vH) filling toward real and complex CDW states has been analyzed using the weak-coupling renormalization group (RG) \cite{Park-prb21, Fernandes-xiv22}, and different CDW phases with LC order have been proposed \cite{Lin-prb21, Feng-sb21, Feng-prb21, Setty-xiv21, Denner-prl21, SZ21, Mertz-npj22, Yang-xiv22, Kontani-xiv22}.
A continuous transition between a topological Chern insulator, analogous to the Haldane model for the quantum anomalous Hall effect \cite{Haldane}, and a topologically trivial CDW insulator was discussed, and subsequent electronic orders due to the residual Fermi surface (FS) sections away from vH filling was speculated in relation to the experiments \cite{Lin-prb21, Lin-prb22}.
On phenomenological grounds, the condensation of the $2a_0 \times 2a_0$ kagom\'e breathing mode with a complex amplitude would give rise to a LC Chern metal with Chern Fermi pockets (CFPs) with concentrated Berry curvature close to but away from the vH filling \cite{SZ21}.
It was pointed out that the circulating LC in the CDW normal state turns into loop-supercurrent in the SC state circulating an emergent vortex-antivortex lattice and gives rise to a roton pair density wave order with the experimentally observed periodicity, with intriguing possibilities of intertwined and vestigial orders including the pseudogap phase and higher-charge superconductivity \cite{SZ21}.
Despite these developments, a microscopic understanding of the LC CDW state in concrete theoretical model realizations on the kagom\'e lattice is still lacking and actively investigated, which is the focus of the current work.

\begin{figure}
\begin{center}
\fig{3.4in}{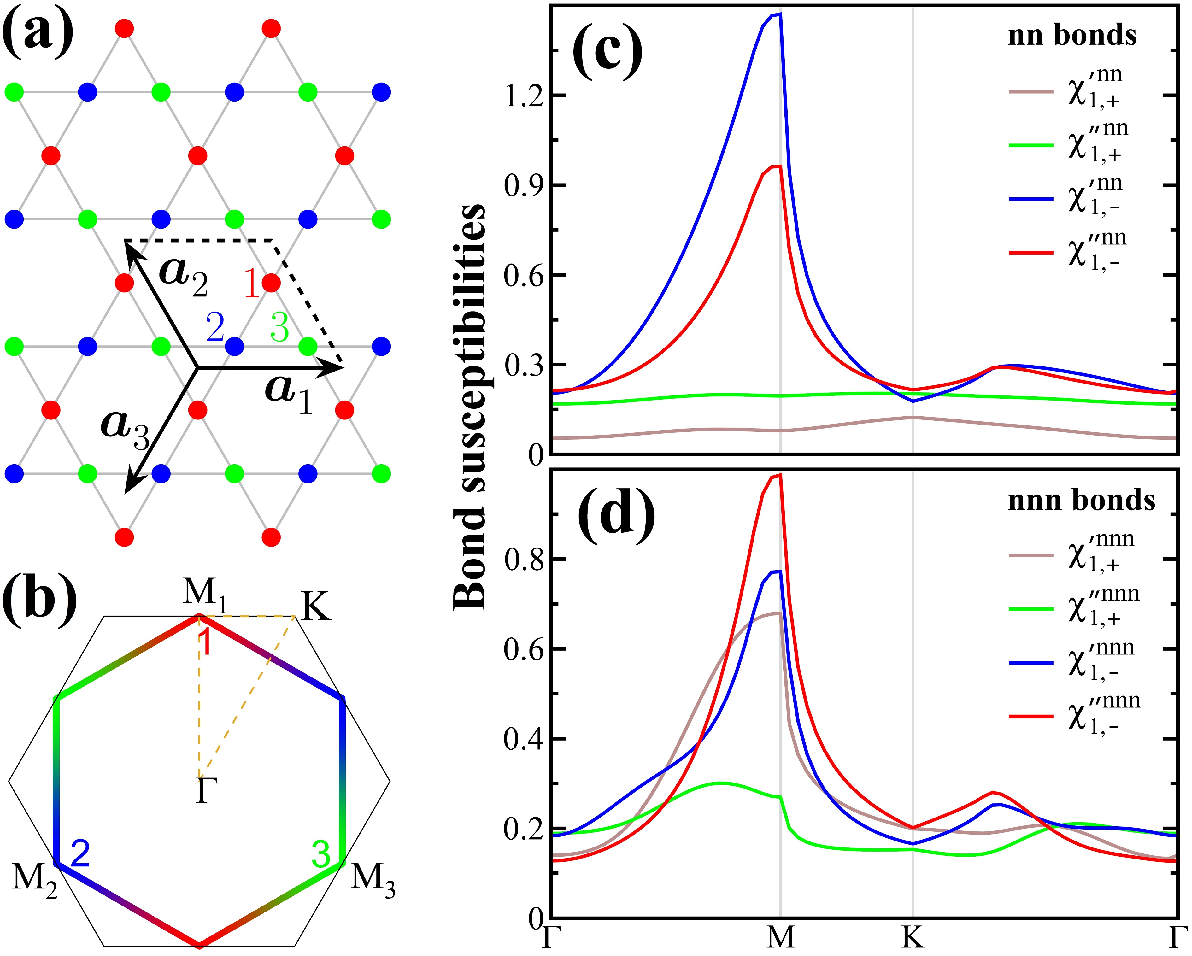}
\caption{(a) Lattice structure of the kagom\'e lattice with \textit{three} sublattices denoted by red (1), blue (2), and green (3) circles.
The two basis lattice vectors are $\ba_1=(1,0)$ and $\ba_2=(-{1\over 2}, {\sqrt{3}\over 2})$, with the third direction follows $\ba_3=-\ba_1-\ba_2$.
The corresponding reciprocal lattice vectors are given by $\bG_1=(0, {4\pi \over \sqrt{3}})$, $\bG_2=(-2\pi, -{2\pi \over \sqrt{3}})$, and $\bG_3=(2\pi, -{2\pi \over \sqrt{3}})$.
(b) Sublattice-resolved FS at vH filling in the hexagonal BZ.
Colors of lines and numbers display the sublattice contents.
The three independent vH points are labeled by $\bM_{1,2,3} ={1\over 2} \bG_{1,2,3}$, respectively.
(c) Bare susceptibilities for the four $\alpha=1$ nn bonds along the high-symmetry path indicated by dashed lines in (b).
A temperature of $k_B T=0.005t$ is applied in the calculation, under which the divergences are reflected by the peaks at $\bM$ point.
(d) Same as (c), but for nnn bonds. }
\label{suscep}
\end{center}
\vskip-0.5cm
\end{figure}

To this end, we focus on the translation symmetry breaking $2a_0\times 2a_0$ CDW and consider single-orbital effective model with electron-electron interactions on the kagom\'e latticed depicted in Fig. \ref{suscep}a, where electrons hop between nearest neighbor (nn) sites and the strengths of interactions are phenomenological input parameters.
At vH filling, the FS of noninteracting electrons passes through the three independent vH points labeled by $\bM_{1,2,3}$ at the zone boundary (Fig.~\ref{suscep}b), that are perfected nested by the $2a_0 \times 2a_0$ 3\bQ-wavevectors $\bQ_\alpha={1\over2}\bG_\alpha$, where $\alpha=1,2,3$ and $\bG_\alpha$ is the reciprocal wavevector of the kagom\'e lattice.
The kagom\'e lattice is special in that the Bloch states at the $\bM_\alpha =\bQ_\alpha$ points are exclusively localized on the $\alpha$-th sublattice due to the sublattice quantum interference effect \cite{Kiesel-prl13, Wu-prl21}, as shown in the sublattice resolved FS in Fig.~\ref{suscep}b.
As a result, the onsite Coulomb interaction $U$, which acts on the same sublattice, is obstructed and ineffective at producing the leading $2a_0 \times 2a_0$ electronic instabilities.
Furthermore, the phenomenological $U$ in the kagom\'e metals is expected to be small since they are found to be nonmagnetic down to very low temperatures \cite{Ortiz-prm19, Ortiz-prm21, Kenney-jpcm21} and their low-energy bands revealed in ARPES agree remarkably well with those predicted in DFT \cite{Ortiz-prl20, Wilson-prx21, Wang-xiv21, Nakayama-xiv21, Hu-xiv21, Kang-xiv21, Luo-nc22}.
Indeed, DFT+DMFT calculations \cite{Zhao-prb21} indicate that these kagom\'e metals are good metals with weak electron correlations, and functional RG studies \cite{Wang-prb13, Kiesel-prl13} suggest that $U$ would favor $\bq=0$ magnetic orders outside the scope of this work.
Therefore, the onsite interaction $U$ is neglected here in the effective model.

Intersite Coulomb interactions connecting different sublattices are expected to play an important role in triggering instabilities towards $2a_0\times 2a_0$ orders.
Extended Hubbard model including the nn Coulomb repulsion $V_1$ on the kagom\'e lattice has been investigated by functional RG \cite{Wang-prb13, Kiesel-prl13}, showing the emergence of translation symmetry breaking bond-ordered CDW and spin density wave at wave vector $\bq=\bQ_{1,2,3}$ when the extended Coulomb $V_1$ dominates.
Despite the emergence of 3\bQ\ real CDW with the SD or ISD bond configurations, no signature of TRS breaking CDW order with LC has been found.

In this work, we argue the importance of next-nn (nnn) interatomic Coulomb repulsion $V_2$ in realizing the 3\bQ\ CDW with spontaneous TRS breaking LC orders near vH filling on the kagom\'e lattice.
The intersite Coulomb interactions have been considered in the original model realization of the LC states for the cuprates \cite{Varma}.
More recently, it has been shown that $V_2$ plays a key role in realizing the correlation-driven topological insulators with orbital currents on the half-filled honeycomb lattice \cite{Raghu-prl08}, and in realizing the ordered LC states in bilayer graphene \cite{Zhu-prb13}.
Moreover, correlation-driven LC states have been unveiled in density-matrix RG studies of the $t$-$V_1$-$V_2$ \cite{Zhu-prb18} and $t$-$V_1$-$V_2$-$V_3$ \cite{Zhu-prl16} models, where $V_3$ is the third-nn Coulomb repulsion, at the filling fraction corresponding to the quadratic and flat band touching point on the kagom\'e lattice.
We are additionally motivated by the physics specific to the kagom\'e lattice close to the vH filling, namely the nesting and sublattice quantum interference effects discussed above. Note that in addition to the interatomic $V_1$ between different nn sublattices, the nnn interaction $V_2$ is also between different sublattices along the three hexagonal directions of the nesting vectors $\bQ_\alpha$, as seen in Fig.~\ref{suscep}a-b, and is therefore free of the sublattice obstruction.
On the other hand, the third-nn $V_3$ is between electrons on the same sublattice, and thus subject to the sublattice obstruction (Fig.~\ref{suscep}a).
We thus keep only $V_1$ and $V_2$ in the truncation of the long-range Coulomb repulsion.

In section II, we calculate the bare susceptibilities of the nn and nnn bonds in the tight-binding model.
The leading divergence at $\bQ_\alpha$ in the nn bond susceptibility turns out to be in the breathing channel \cite{SZ21} of the real part of the bond operator, whereas in the nnn bond susceptibility, it happens in the breathing channel of the imaginary part of the nnn bond operator.
This suggests that, while the nn Coulomb repulsion $V_1$ favors real bond CDW order, the nnn Coulomb repulsion $V_2$ favors imaginary bond orders, pointing to the importance of $V_2$ in realizing LC order.

In section III, we introduce the minimal $t$-$V_1$-$V_2$ effective model.
To capture the basic physics of the leading instability at the vH filling and reveal the nature of the competitive bond CDW states, we project the inter-site $V_1$ and $V_2$ interaction terms onto the vH states.
This puts the physical states of the electrons at the $\bM_\alpha$ points.
We find an exact result that the projected inter-sublattice Coulomb interactions $V_1$ and $V_2$ couple precisely the antisymmetry bond, corresponding to the nn and nnn breathing modes \cite{SZ21}, respectively.
The weak-coupling instability at vH filling is thus completely determined by the aforementioned bond susceptibilities, which allow us to show that while $V_1$ drives a real bond CDW order, $V_2$ drives a competing imaginary bond CDW order corresponding to a LC Chern insulator.

In section IV, we perform a self-consistent mean-field calculation for the $t$-$V_1$-$V_2$ model and obtain the phase diagram that reveals the orbital Chern insulator phases of different LC patterns, separated from the topological trivial ISD phase by a first order phase transition.
We further demonstrate that electron doping away from vH filling helps the stabilization of the LC states, giving rise to the conjectured LC Chern metal with CFPs carrying large Berry curvature and orbital magnetic moment \cite{SZ21}.
The summary and discussions are provided in section V in connection to the experimental evidence for the TRS breaking charge ordered states in the kagom\'e superconductors.

\section{Tight-binding model and bond susceptibilities.}
The single-orbital tight-binding model with nn hopping $t$ on the kagom\'e lattice is depicted in Fig. \ref{suscep}a.
The locations of the three sublattices in the unit cell at \br\ are given by $\br_1=\br -{1\over 2}\ba_3$, $\br_2 =\br$, and $\br_3= \br+ {1\over 2}\ba_1$, with the lattice constant $a_0\equiv 1$.
Along each $\ba_\alpha$ direction, there are two independent nn bonds within a unit cell, which define the symmetric ($+$) and antisymmetric ($-$) bond operators \cite{SZ21},
\begin{equation}
\hat\chiup^{\rm nn}_{\alpha,\pm}(\br) = c^\dagger_{\beta\br} c_{\gamma \br} \pm c^\dagger_{\beta\br} c_{\gamma \br-\ba_\gamma},
\label{nn-bond}
\end{equation}
where $c^\dagger_{\alpha \br}$ creates an electron on sublattice $\alpha$ in unit cell $\br$.
The sublattice indices run over $(\alpha,\beta,\gamma)=$ (1, 2, 3), (2, 3, 1), (3, 1, 2) and spin indices are left implicit. The unique sublattice geometry of the kagom\'e lattice is clearly revealed by rewriting the tight-binding model as
\begin{equation}
H_\text{tb}= -t \sum_{\alpha,\br} \left[ \hat\chiup^{\rm nn}_{\alpha,+}(\br) + h.c. \right]
-\mu \sum_{\alpha,\br} c_{\alpha\br}^\dagger c_{\alpha\br},
\label{htb}
\end{equation}
where $\mu$ is the chemical potential.
To study the instabilities toward the real and imaginary CDW in the in-phase and the out-of-phase (breathing) bond channels separately, we divide the symmetric and antisymmetric bonds into their real and imaginary components,
\begin{equation}
\hat\chiup^{\rm\prime nn}_{\alpha,\pm} = {1\over 2} \left[ \hat\chiup^{\rm nn}_{\alpha,\pm} +  (\hat\chiup^{\rm nn}_{\alpha,\pm})^\dagger \right], \quad
\hat\chiup^{\rm\prime\prime nn}_{\alpha,\pm} = {1\over 2 i} \left[\hat\chiup^{\rm nn}_{\alpha,\pm} -  (\hat\chiup^{\rm nn}_{\alpha,\pm} )^\dagger \right],
\end{equation}
and compute the corresponding bare susceptibilities for the tight-binding noninteracting electrons
\begin{equation}
\Pi_{ \hat O} (\bq) = \lim_{\omega =0} i \int^\infty_0 dt e^{i\omega t} \left \langle \left[ {\hat O} (\bq, t), {\hat O}^\dagger (-\bq, 0) \right] \right\rangle,
\end{equation}
with ${\hat O}$ being one of the four bond operators in each $\ba_\alpha$ direction, $\hat\chiup^{\rm\prime nn}_{\alpha,\pm}$ and $\hat\chiup^{\rm\prime\prime nn}_{\alpha,\pm}$.
A temperature of $k_B T = 0.005t$ is applied in the calculation to avoid the logarithmic divergences tied to the vH singularities.

At band filling $n_\text{vH} = 5/12$, the sublattice-resolved FS is the hexagon connecting the vH singularities at the $\bM$ points of the BZ, as shown in Fig. \ref{suscep}b.
The three vH points are connected to each other by the $2\times2$ wavevectors $\bQ_\alpha=\bM_\alpha$.
Since the Bloch states at these three vH points are localized on different sublattices, the instability towards $2\times2$ onsite charge density order on the same sublattice is significantly suppressed, and the most likely $2\times2$ order is off-site bond order that connect two sites belonging to two different sublattices.
The calculated bare susceptibilities of the four nn bonds in the $\alpha=1$ direction, i.e., $ \hat\chiup^{\rm\prime nn}_{1,\pm}$ and $\hat\chiup^{\rm\prime\prime nn}_{1,\pm}$, are plotted along the high-symmetry path in the BZ and compared directly in Fig. \ref{suscep}c.
The susceptibilities in the $\alpha\neq 1$ directions can be obtained from $\hat\chiup^{\rm\prime nn}_{1,\pm}$ and $\hat\chiup^{\rm\prime\prime nn}_{1,\pm}$ by $C_3$ rotations.
It is clear that the susceptibilities of the symmetric nn bonds are much weaker than those of the antisymmetric nn bonds in the breathing channel with the divergence reflected by the peaks located at $\bQ_1$, the wavevector connecting the two vH points containing the sublattice 2 and 3, as shown in Fig. \ref{suscep}b.
The leading divergence is in the susceptibility of the real antisymmetric bonds, pointing to leading instability toward a $2\times2$  real CDW driven by nn interaction $V_1$, consistent with previous investigations \cite{Wang-prb13, Kiesel-prl13}.

The susceptibilities of the nnn bonds are intriguing.
Just as for the nn bonds, there are two independent nnn bonds in a unit cell along the direction \textit{perpendicular} to each basis vector $\ba_\alpha$ (i.e., along $\bQ_\alpha$ direction), connecting two sites of different sublattices, as shown in Fig.~\ref{suscep}a.
Define symmetric ($+$) and antisymmetric ($-$) nnn bonds, and divide them into real and imaginary components,
\begin{equation}
\hat\chiup^{\rm nnn}_{\alpha,\pm} (\br) = c^\dagger_{\beta\br} c_{\gamma \br+\ba_\beta} \pm c^\dagger_{\beta\br} c_{\gamma \br+\ba_\gamma} =\hat\chiup^{\rm \prime nnn}_{\alpha,\pm} (\br) +i\hat\chiup^{\rm\prime\prime nnn}_{\alpha,\pm} (\br).
\label{nnn-bond}
\end{equation}
The calculated bare susceptibilities of the four nnn bonds in the $\alpha=1$ direction, i.e., $ \hat\chiup^{\rm\prime nnn}_{1,\pm}$ and $\hat\chiup^{\rm\prime\prime nnn}_{1,\pm}$, are plotted in Fig. \ref{suscep}d.
In contrast to the nn bonds, the imaginary nnn bond in the antisymmetric breathing channel $\hat\chiup^{\rm\prime\prime nnn}_{1,\pm}$ has now the leading divergent susceptibility, while the real counterpart $\hat\chiup^{\rm\prime nnn}_{1,\pm}$ is subleading.
As a result, we expect the nnn Coulomb interaction $V_2$ on kagom\'e lattice to play a crucial role in driving an instability toward the spontaneous TRS breaking $2\times2$ CDW with LC order.
This motivates us to study the concrete $t$-$V_1$-$V_2$ model.

\section{Interatomic interactions and projection to van Hove modes}
\subsection{The $t$-$V_1$-$V_2$ model}
The single-orbital $t$-$V_1$-$V_2$ model is given by
\begin{equation}
H = - t \sum_{\langle i, j \rangle \sigma} (c^\dagger_{i\sigma} c_{j \sigma} + h.c.) +V_1 \sum_{\langle i, j \rangle} \hat{n}_i \hat{n}_j +V_2 \sum_{\llangle i, j \rrangle} \hat{n}_i \hat{n}_j,
\label{tv1v2}
\end{equation}
where $c^\dagger_{i \sigma}$ creates a spin-$\sigma$ electron on site $i$, and the density operator $\hat{n}_i=\sum_\sigma c^\dagger_{i\sigma} c_{i\sigma}$.
Hereafter, we set the nn hopping $t\equiv1$ to be the energy unit.
To bring out the sublatice structures on the kagom\'e lattice, we write
\begin{equation}
H=H_{\rm tb} + H_{V_1} + H_{V_2},
\end{equation}
where $H_{\rm tb}$ is the tight-binding part given in Eq.~(\ref{htb}) and
\begin{align}
H_{V_1}=&V_1\sum_{\alpha,\br}(\hat n_{\beta\br}\hat n_{\gamma\br}
+\hat n_{\beta\br}\hat n_{\gamma\br-{\ba_\alpha}})
\label{hv1} \\
H_{V_2}=&V_2\sum_{\alpha,\br}(\hat n_{\beta\br}\hat n_{\gamma
\br+\ba_{\beta}} +\hat n_{\beta\br}\hat n_{\gamma\br +\ba_{\gamma}}).
\label{hv2}
\end{align}
Expanding the electron operator on the $\alpha$ sublattice located in the $\br$ unit cell in terms of the complete set of momenta in the first BZ,
\begin{equation}
c_{\alpha\br} =\sum_{\bk\in1BZ} e^{-i\bk\cdot\br} c_{\alpha\bk},\quad
c_{\alpha\br}^\dagger =\sum_{\bk\in1BZ} e^{i\bk\cdot\br} c_{\alpha\bk}^\dagger,
\label{ft}
\end{equation}
we obtain the Fourier transform of the inter-site interactions,
\begin{equation}
H_{V_{1,2}}
=-\sum_{\alpha\bq}\sum_{\bk\bk^\prime}V_{1,2}^\alpha(\bk,\bk^\prime,\bq)
c_{\beta\bk}^\dagger c_{\gamma\bk+\bq}c_{\gamma\bk^\prime}^\dagger c_{\beta\bk^\prime-\bq},
\label{v12k}
\end{equation}
where the corresponding form factors for the nn $V_1$ and nnn ${V_2}$ Coulomb repulsions are given by
\begin{align}
V_1^\alpha(\bk,\bk^\prime,\bq)=& V_1[1 +e^{i(\bk-\bk^\prime+\bq) \cdot \ba_\alpha}], \label{f1} \\
V_2^\alpha(\bk,\bk^\prime,\bq)=& V_2 [e^{-i(\bk-\bk^\prime+\bq)\cdot \ba_{\beta}}+ e^{-i(\bk-\bk^\prime+\bq)\cdot \ba_{\gamma}}].
\label{f2}
\end{align}

\subsection{Projected interactions and van Hove modes}
At vH filling, the tight-binding FS is the hexagon with corners located at the vH points $\bM_\alpha$ where the density of states diverges (Fig.~\ref{suscep}b).
To study the leading instability of the electronic states, we project the physical states onto these vH modes under the projection operator $\hat{P}=\sum_\alpha \vert \bM_\alpha\rangle \langle \bM_\alpha\vert$.
Since these states are localized on independent sublattices, the projection is accomplished by putting the momenta of the electron operators in Eq. (\ref{v12k}) on the vH points of the corresponding sublattices, $\bk=\bk^\prime-\bq=\bM_{\beta}$ and $\bk+\bq =\bk^\prime=\bM_{\gamma}$, which amounts to $\bk=\bM_{\beta}$, $\bk^\prime=\bM_{\gamma}$, and $\bq=\bM_{\alpha}$.
To evaluate the form factors in Eqs~(\ref{f1}-\ref{f2}), we note that $\bM_\alpha \cdot \ba_\beta = {1\over2} \bG_\alpha \cdot \ba_\beta = \pi(1-\delta_{\alpha \beta})$ (mod $2\pi$).
Thus, the projected interactions read
\begin{equation}
H_{V_{1,2}}^{\rm vH}=-2V_{1,2} \sum_\alpha
c_{\beta \bM_\beta}^\dagger c_{\gamma \bM_\gamma} c_{\gamma \bM_\gamma}^\dagger c_{\beta \bM_\beta}.
\label{hv12vh}
\end{equation}
Clearly, the extended Coulomb repulsions $V_1$ and $V_2$ couple directly to the vH modes in a similar form.

It is instructive to write the projected interactions in terms of the bond operators.
Since the nn and nnn bond operators in momentum space are given by
\begin{align}
\hat{\chiup}^{\rm nn}_{\alpha,\pm} (\bq) &=\sum_\bk \left[ 1\pm e^{i(\bk+\bq) \cdot \ba_\alpha} \right] c^\dagger_{\beta\bk} c_{\gamma \bk+\bq}, \\
\hat{\chiup}^{\rm nnn}_{\alpha,\pm} (\bq) &=\sum_\bk \left[ e^{-i(\bk+\bq) \cdot \ba_\beta} \pm e^{-i(\bk+\bq) \cdot \ba_\gamma} \right] c^\dagger_{\beta\bk} c_{\gamma \bk+\bq},
\end{align}
projecting the fermions onto the vH modes yields
\begin{align}
\hat{\chiup}^{\rm nn}_{\alpha,+} (\bM_\alpha) & = \hat{\chiup}^{\rm nnn}_{\alpha,+} (\bM_\alpha) =0, \\
\hat{\chiup}^{\rm nn}_{\alpha,-} (\bM_\alpha) & = -\hat{\chiup}^{\rm nnn}_{\alpha,-} (\bM_\alpha) = 2c^\dagger_{\beta \bM_\beta} c_{\gamma \bM_\gamma}.
\end{align}
Consequently, the intersite interactions in the projected vH manifold in Eq.~(\ref{hv12vh}) become
\begin{equation}
H_{V_1}^{\rm vH}=-{V_1 \over 2} \sum_\alpha \left| {\hat\chiup}_{\alpha,-}^{\rm nn}(\bM_\alpha) \right|^2, \quad
H_{V_2}^{\rm vH}=-{V_2\over 2} \sum_\alpha \left| {\hat\chiup}_{\alpha,-}^{\rm nnn}(\bM_\alpha) \right|^2. \nonumber
\end{equation}
These results demonstrate explicitly that the extended Coulomb interactions drive the breathing instability of the kagom\'e lattice represented by the antisymmetric bonds \cite{SZ21}. Combined with the bare susceptibilities calculated for the antisymmetry bonds shown in Fig.~\ref{suscep}c-d, it is clear that there is a weak-coupling instability toward a real CDW due to nn $V_1$, and one toward an imaginary CDW with LC due to nnn $V_2$, respectively.
When both $V_1$ and $V_2$ are present, we find that the Landau free energy analysis \cite{Park-prb21} predicts a transition in the unstable vH modes toward an imaginary CDW with spontaneous TRS breaking when
\begin{equation}
{V_2 \over V_1} > {\Pi_{\alpha,-}^{\prime\rm nn}(\bM_\alpha) - \Pi_{\alpha,-}^{\prime\prime \rm nn}(\bM_\alpha) \over \Pi_{\alpha,-}^{\prime\prime \rm nnn}(\bM_\alpha) - \Pi_{\alpha,-}^{\prime \rm nnn}(\bM_\alpha)}.
\end{equation}
The value of this critical ratio can be estimated using the corresponding susceptibilities of the nn and nnn antisymmetric bonds in Fig.~\ref{suscep}c-d, leading to $V_2/V_1 > (1.47 - 0.96)/( 0.99 - 0.77) \simeq 2.36$.

\section{Mean-field theory and phase diagram}
\subsection{Mean-field theory}
Next, we go beyond the weak-coupling analysis and study the phase structure of the $t$-$V_1$-$V_2$ model in Eq.~(\ref{tv1v2}) as a function of the correlations $V_1$ and $V_2$ in a self-consistent mean-field theory.
Decoupling the Coulomb repulsions $V_1$ and $V_2$ in terms of the nn and nnn bonds, $\hat{\chiup}_{ij} =c^\dagger_{i\uparrow} c_{j\uparrow} +c^\dagger_{i\downarrow} c_{j\downarrow}$, we obtained the mean-field Hamiltonian in real space,
\begin{align}
H_\text{MF} = -&\sum_{\langle i,j \rangle} \left[ (t+V_1\chiup^*_{ij}) \hat{\chiup}_{ij} +h.c. -V_1 |\chiup_{ij}|^2 \right] \nonumber \\
-&\sum_{\llangle i,j \rrangle} \left( V_2 \chiup^*_{ij} \hat{\chiup}_{ij} +h.c. -V_2 |\chiup_{ij}|^2 \right).
\label{hmf}
\end{align}
The order parameters of the nn and nnn bonds $\chiup_{ij}=\langle \hat{\chiup}_{ij} \rangle$ are to be determined self-consistently.

A few remarks on the mean-field theory are in order.
First of all,  we consider the $t$-$V_1$-$V_2$ model as an effective model for capturing the correlation effects beyond the DFT calculations, and as such, the direct Hatree terms from $V_1$ and $V_2$ depending on the total density $n_i=\langle \hat{n}_i \rangle$ are neglected to avoid double-counting \cite{Jiang-prb16}, similar to the LDA+$U$+$V$ approach \cite{Campo-JPCM10, Belozerov-prb12}.
Moreover, note that the real uniform components of the bond order parameters $\chiup_{ij}$ in Eq.~(\ref{hmf}) induced by the Coulomb repulsions retain all the symmetries under the point group $D_6$ of the kagom\'e lattice.
They merely introduce corrections to the nn and nnn hopping integrals that renormalize the bandwidth and modify or even remove the vH singularity at the $\bM$ points.
Since the vH singularity at the $\bM$ points have been established by ARPES in all \avs, we adapt a procedure where all the symmetry invariant corrections to the real uniform order parameter are subtracted in the interactions, such that the vH points and the band structure are maintained without adjusting the bare band parameters.
Similar subtraction schemes have been used successfully in the Hartree-Fock mean-field calculations of the Coulomb interaction in twisted bilayer graphene \cite{Zhang-prb20, Xie-prl20, Ashvin-prx20, Ashvin-prr21}.
Implementing this subtraction scheme, the correlation effects in the self-consistent mean-field theory will only generate spontaneous symmetry-breaking states.
We have verified that the subtraction only quantitatively shift the phase boundary in the phase diagram without changing our main conclusions on the emergence of the LC state.

\begin{figure}
\begin{center}
\fig{3.4in}{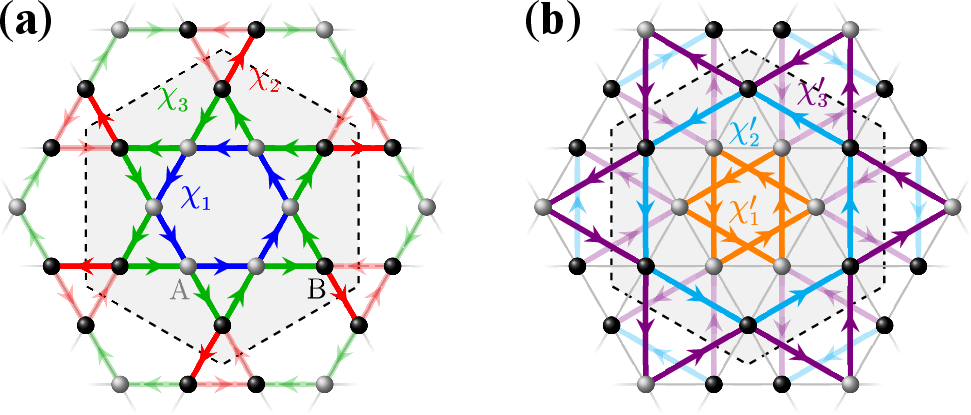}
\caption{Schematic representation of (a) nn and (b) nnn bonds in the $2\times 2$ unit cell enclosed by dashed lines.
$C_6$ symmetry and continuity condition dictate equality of sites, nn and nnn bonds, as indicated by the colors of the spheres and bonds.
Arrows denote the flowing directions of currents.
The colors of bonds replicated by $2\times 2$ translation are weakened for a better visualization.}
\label{pattern}
\end{center}
\vskip-0.65cm
\end{figure}

\subsection{Bond order and LC patterns}
To capture the physics in the vicinity of vH filling, we consider the quantum states described by the Hamiltonian in Eq. (\ref{hmf}) to be periodic with an enlarged $2 \times 2$ unit cell, enclosed by black dashed lines in Fig. \ref{pattern}.
The enlarged unit cell contains 12 sites, 24 nn bonds, and 24 nnn bonds.
A completely unbiased solution of the $2 \times 2$ superlattice thus involves determining the 48 independent bonds self-consistently.
In this work, we focus on states with symmetries described by the point group $C_6$, which is a subgroup of $D_6$.
This includes the LC states that break certain reflection symmetries spontaneously.
The 6 sites on the inner/outter apexes of the green stars in Fig. \ref{pattern}a can be mapped onto each other by the point group symmetry operation of $C_6$ and, consequently, belong to the same class of sites with identical properties.
These two classes of sites are labeled as $A$ and $B$, respectively, and denoted by grey and black spheres in Fig. \ref{pattern}.
For the nn and nnn bonds, we note that, in addition to the constraints imposed by $C_6$ symmetry, currents associated with their imaginary parts must satisfy the continuity condition for charges to be conserved.
As a result, the 24 nn bonds are reduced to three distinct classes as illustrated in Fig. \ref{pattern}a.
Specifically, the 6 blue bonds forming the inner hexagon connect the $A$ sites and are labeled by $\chiup_1$ in Fig. \ref{pattern}a; the 6 red bonds ($\chiup_2$) of the outer two triangles link the $B$ sites; and the 12 green bonds ($\chiup_3$) of the star bridge the $A$ and $B$ sites.
Similarly, the 24 nnn bonds are also reduced to three independent classes of equivalent ones, denoted by the colors of bonds and labeled as $\chiup'_{1,2,3}$ in Fig. \ref{pattern}b.
We note that, in these $C_6$ symmetric states, the average density on the four nn sites and that on the four nnn sites of any lattice site equal to the average density of the system.
As a result, the Hartree term, even if considered, would only introduce effectively a global chemical potential term to the Hamiltonian and thus physically irrelevant.

In Fig. \ref{pattern}, the anticlockwise winding arrows on each class of bonds signify the forward flowing direction of the currents on these bonds.
They define the current direction associated with a positive $\text{Im}(\chiup_{1,2,3})$ or $\text{Im}(\chiup'_{1,2,3})$.
The physical current lives on the nn bonds and the current on the nnn bonds is associated with the quantum fluctuations of the charge density.
The current direction on each class of bonds can be reversed without violating any constraint imposed by $C_6$ symmetry or the continuity condition.
Note that two states with all currents flowing in opposite directions can be mapped to each other by reflection, and share identical energy and quasiparticle band structure.
Furthermore, as we shall show later, physical properties, topological ones in particular, are controlled by the current pattern on the nn bonds.
Therefore, we fix the current direction on $\chiup_1$ bonds to be anticlockwise, and use the current pattern on the three classes of the nn bonds to define the LC states.
Consequently, there are in total four possible LC states:
LC$_1$ with currents flowing anticlockwise in all three classes of nn bonds (Fig. \ref{pattern}a), LC$_2$ with currents on $\chiup_2$ bonds flipped, LC$_3$ with currents on $\chiup_3$ bonds reversed, and LC$_4$ with currents on both $\chiup_2$ and $\chiup_3$ bonds over-turned.
We will show that all these LC states appear on the phase diagram spanned by $V_1$ and $V_2$ in the self-consistent mean-field theory.

In order to obtain all the possible states, we use different initial conditions for solving the self-consistency equations numerically.
Besides the four LC states mentioned above, we also encounter converged solutions of two real CDW states, SD and ISD, where all bonds are real and the LC is absent.
In these real CDW states, $\chiup_1$ and $\chiup_2$ are close in value, which is weaker (stronger) than $\chiup_3$ in the SD (ISD) configuration with the stronger bonds forming the stars (hexagon and triangles).
Moreover, the LC states we found have complex bonds $\chiup_{1,2,3}$ and $\chiup^\prime_{1,2,3}$ and are thus complex bond CDW states.
While the real part of these complex bonds may deviate from SD or ISD configurations, we will denote these states according to their current patterns as LC$_{1,\dots,4}$, unless otherwise notes.
When multiple converged states exist for a given set of parameters, $V=(V_1, V_2)$, we compare the state energies to determine the true ground state.

\begin{figure}
\begin{center}
\fig{3.4in}{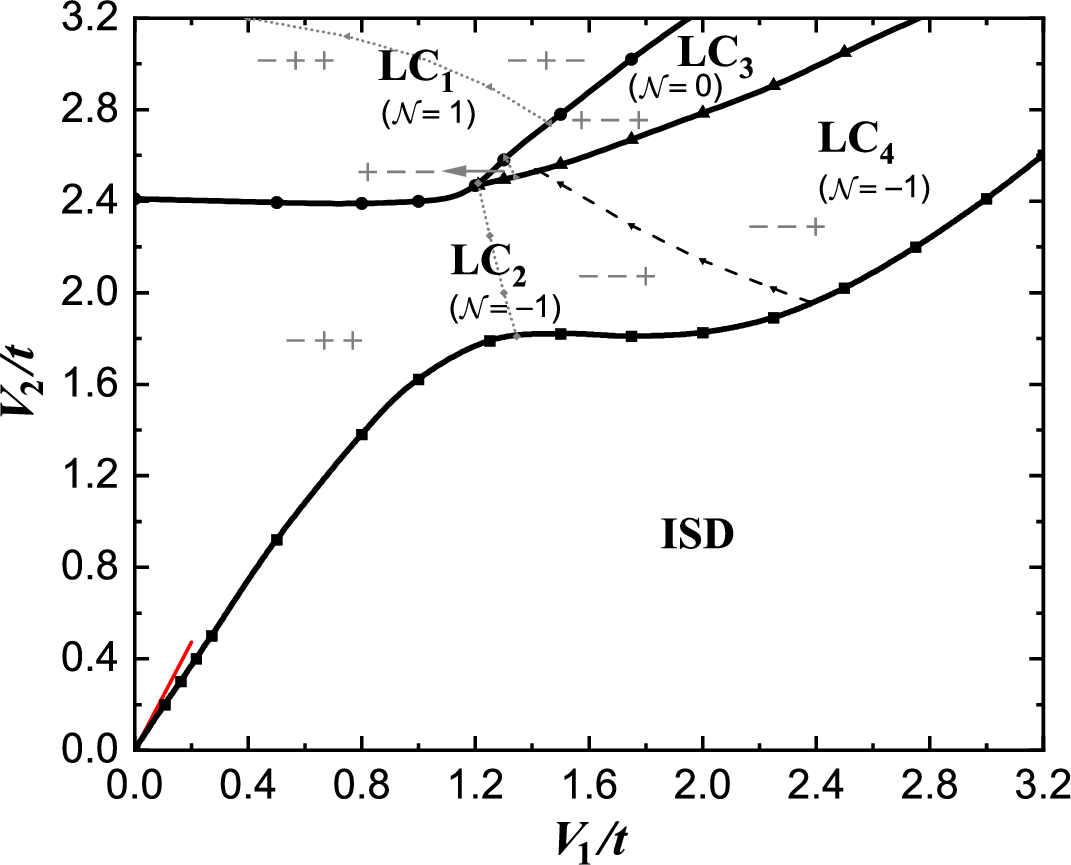}
\caption{Ground state phase diagram of the kagom\'e lattice $t$-$V_1$-$V_2$ model at vH filling.
Solid thick lines denote phase boundaries of first-order transitions among topologically trivial ISD and topologically nontrivial LC states, with the total Chern number ${\cal N}$ displayed for LC states.
LC$_2$ and LC$_4$ separated by dashed thin line are of the same phase with identical topology.
The red thin line of slope $\sim 2.36$ denote the boundary between real and imaginary CDW estimated by the weak-coupling analysis in section III.
Additional dotted thin lines are additional boundaries separating different current patterns on nnn bonds indicated by grey symbols "$\pm\pm\pm$".}
\label{diagram}
\end{center}
\vskip-0.65cm
\end{figure}

\subsection{Phase diagram and transitions at vH filling}
\subsubsection{Phase structure of ISD and LC states}
We first study the zero-temperature phase structure and the formation of the LC states at vH filling with $n=n_\text{vH} = 5/12$.
The ground state phase diagram is presented in Fig. \ref{diagram} as a function of nn $V_1$ and nnn $V_2$.
In the absence of nnn Coulomb repulsion, $V_2=0$, only two states can be self-consistently obtained, ISD and LC$_2$, with the ISD state lower in energy in the entire parameter regime we explored.
Thus, LC$_2$ is a metastable state with higher energy in this case.
In LC$_2$ at $V_2=0$, the real parts of $\chiup_1$ and $\chiup_2$ are close in value and much larger than $\text{Re}(\chiup_3)$, forming a SD pattern.
Switching on the nnn Coulomb repulsion $V_2$, we find that various LC states can be stabilized as the ground state when $V_2$ is sufficiently large and the ISD gives way to the LC states via a first-order transition.
Overall, the currents on $\chiup_3$ bonds tend to flow anticlockwise at small $V_1$ and clockwise at large $V_1$, while the currents on $\chiup_2$ bonds prefer to flow clockwise at small $V_2$ and anticlockwise at large $V_2$, producing the rich phase diagram shown in Fig. \ref{diagram}, where LC$_2$, the LC state studied phenomenologically in Ref.~\cite{SZ21}, occupies the phase regime for small and moderate $V_1$ and $V_2$.

\begin{figure*}
\begin{center}
\emph{\emph{}}\fig{7in}{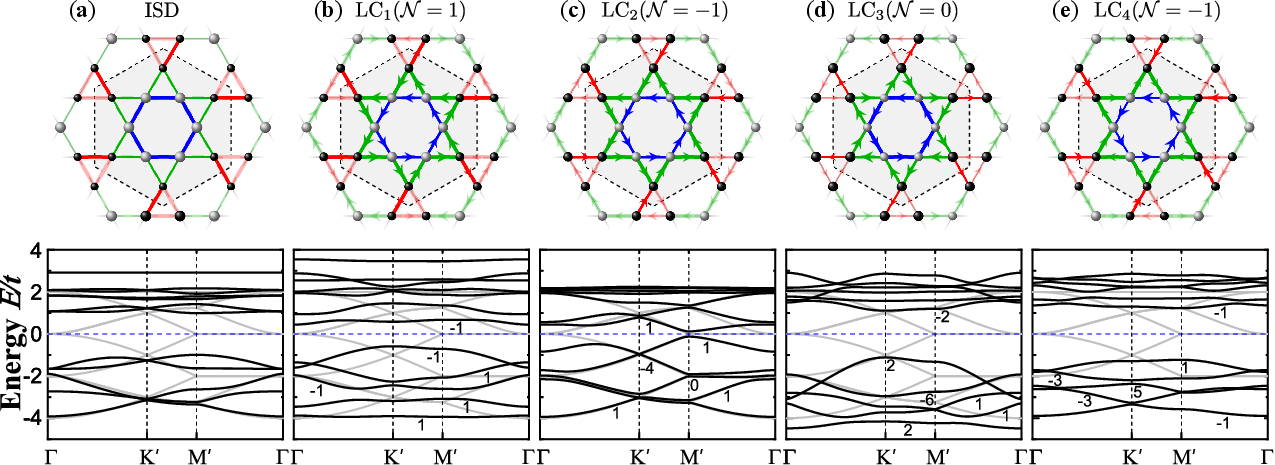}
\caption{Schematic representation (upper panels) and band dispersion along high-symmetry path in the reduced BZ (lower panels) for the five ground states at vH filling.
(a) ISD at $V=(2, 1)$,
(b) LC$_1$ at $V=(0.5, 2.5)$,
(c) LC$_2$ at $V=(0.8, 1.6)$,
(d) LC$_3$ at $V=(2, 3)$,
(e) LC$_4$ at $V=(2, 2.5)$.
Size of spheres denotes the local electron densities, width of bonds and size of arrows indicate the values of real and imaginary part of nn bonds, respectively.
Numbers in lower panels denote the Chern numbers of corresponding bands.
The tight-binding dispersions folded into the reduced BZ are also plotted in grey lines for comparison.}
\label{states}
\end{center}
\end{figure*}

The phase diagram (Fig.~\ref{diagram}) consists of five ground states, an ISD and four LC states.
Their schematic illustration at representative values of $V=(V_1,V_2)$ and the corresponding quasiparticle dispersions are displayed in Fig.~\ref{states}, with the self-consistently determined nn and nnn bond expectation values listed in Table I.
Because of the instabilities at vH filling, ground states in the entire phase diagram, except for the noninteracting point at the origin $V_{1,2}=0$, exhibit spontaneous symmetry breaking and are insulating states, which can be seen in the quasiparticle dispersions.
The ISD state breaks only the lattice translation symmetry and is topologically trivial, while the LC states break TRS additionally and are topologically nontrivial on the kagom\'e lattice.

\begin{table*}[htb]
\caption{The charge disproportionation $\delta$, the nn bonds $\chiup_{1,2,3}$, and the nnn bonds $\chiup'_{1,2,3}$ of the ground states shown in Fig. \ref{states} at vH filling with $n=n_\text{vH} =5/12$ and those shown in Fig. \ref{awayvH} away from vH filling with $n=n_\text{ed}=5.064/12$.}
\label{values}
\begin{ruledtabular}
\begin{tabular}{c|c|c|c|c|c|c|c|c}
 & $n$ & $\delta$ & $\chiup_1$ & $\chiup_2$ & $\chiup_3$ & $\chiup'_1$ & $\chiup'_2$ & $\chiup'_3$ \\
\hline
\text{ISD at} & $n_\text{vH}$ & 0.115 & 0.631 & 0.636 & 0.181 & 0.007 & 0.025 & $-0.071$ \\
\cline{2-9}
$V=(2,1)$ & $n_\text{ed}$ & 0.110 & 0.622 & 0.628 & 0.194 & 0.002 & 0.022 & $-0.078$ \\
\hline
{LC$_1$ at} & $n_\text{vH}$ & 0.053 & $0.255+0.045i$ & $0.411+0.015i$ & $0.381+0.060i$ & $-0.115-0.450i$ & $0.037+0.294i$ & $-0.004+0.050i$ \\
\cline{2-9}
$V=(0.5,2.5)$ & $n_\text{ed}$ & 0.053 & $0.265+0.048i$ & $0.424+0.015i$ & $0.367+0.059i$ & $-0.101-0.454i$ & $0.051+0.293i$ & $-0.022+0.048i$ \\
\hline
{LC$_2$ at} & $n_\text{vH}$ & 0.007 & $0.378+0.069i$ & $0.375-0.069i$ & $0.490+0.064i$ & $-0.060-0.109i$ & $-0.068+0.075i$ & $-0.001+0.095i$ \\
\cline{2-9}
$V=(0.8,1.6)$ & $n_\text{ed}$ & 0.005 & $0.384+0.066i$ & $0.379-0.069i$ & $0.485+0.066i$ & $-0.062-0.112i$ & $-0.079+0.075i$ & $-0.004+0.093i$ \\
\hline
{LC$_3$ at} & $n_\text{vH}$ & $-0.076$ & $0.453+0.161i$ & $0.162+0.047i$ & $0.348-0.205i$ & $0.087+0.235i$ & $-0.279-0.421i$ & $0.045+0.020i$ \\
\cline{2-9}
$V=(2,3)$ & $n_\text{ed}$ & $-0.080$ & $0.456+0.155i$ & $0.171+0.059i$ & $0.353-0.201i$ & $0.083+0.230i$ & $-0.290-0.398i$ & $0.043+0.009i$ \\
\hline
{LC$_4$ at} & $n_\text{vH}$ & 0.045 & $0.179+0.334i$ & $0.359-0.182i$ & $0.466-0.022i$ & $-0.181-0.114i$ & $0.053-0.127i$ & $0.018+0.235i$ \\
\cline{2-9}
$V=(2,2.5)$ & $n_\text{ed}$ & 0.048 & $0.185+0.321i$ & $0.370-0.177i$ & $0.461-0.020i$ & $-0.189-0.131i$ & $0.058-0.117i$ & $0.010+0.231i$
\end{tabular}
\end{ruledtabular}
\end{table*}

\subsubsection{Topological properties of the LC states}
Fig. \ref{states} shows that each isolated quasiparticle band in LC states acquires an integer topological Chern number.
It is important to note that, while the Chern number of each band may change its value as the interaction parameters $(V_1, V_2)$ change in a particular LC regime in the ground state phase diagram, the total Chern number of all occupied bands ${\cal N}$ remains unchanged and therefore serves as the integer topological invariant of the LC state.
Explicitly, we find ${\cal N} =1$, $-1$, 0, and $-1$, respectively, for LC$_1$, LC$_2$, LC$_3$, and LC$_4$.
The ${\cal N} = \pm 2$ LC states predicted by statistical learning \cite{Mertz-npj22} do not appear in the mean-field phase diagram of the $t$-$V_1$-$V_2$ model.
Interestingly, LC$_2$ and LC$_4$ separated by dashed thin line in Fig. \ref{diagram} have the same ${\cal N}$ and will be shown below to be in the same phase, whereas all other boundaries between two states with different total Chern numbers ${\cal N}$ are first-order phase transitions, as indicated by solid thick lines in the phase diagram (Fig. \ref{diagram}).
The insulating LC states with nonzero total Chern numbers are thus orbital Chern insulators, since the TRS is broken by the orbital LC order \cite{Feng-sb21,Lin-prb21,SZ21}. Moreover, since the LC order has a nonzero momentum, the sum of the staggered flux around each irreducible plaquette in the $2\times2$ unit cell is strictly zero, similar to an antiferromagnetic (AF) ordered state. Therefore, the LC states can be viewed as orbital antiferromagnets \cite{SZ21}. The states LC$_{1,2,4}$ have nonzero ${\cal N}$ and are therefore orbital AF Chern insulators, i.e. orbital analogs of the spin AF Chern insulators \cite{Jiang-prl18}.

The current pattern on the nnn bonds in different LC states in the phase diagram is denoted by grey symbols ``$\pm\pm\pm$" in Fig. \ref{diagram}.
They refer to the directions of the currents on the nnn bonds, i.e. the sign of ${\rm Im}(\chiup_{1,2,3}^\prime)$, with the``+'' direction defined as in Fig.~\ref{pattern}b.
There exist two kinds of current patterns on the  nnn bonds in the phases of LC$_{1,2,3}$, which are separated by dotted thin lines.
We stress that the total Chern number ${\cal N}$ does not change its value as the currents on nnn bonds change their patterns continuously across the dotted thin lines in Fig.~\ref{diagram}, which implies ${\cal N}$ is solely determined by the current pattern on the nn bonds, supporting our choice of using the latter alone to define these TRS broken states.
However, as discussed below, the imaginary part of the nnn bonds can be quite large, sometimes even larger than that of the nn bonds.
It is thus interesting to figure out the effects of the nnn bond on the electronic structure, band topology, and other physical quantities.

\subsubsection{Transitions between ISD and LC states}
\begin{figure}
\begin{center}
\fig{3.4in}{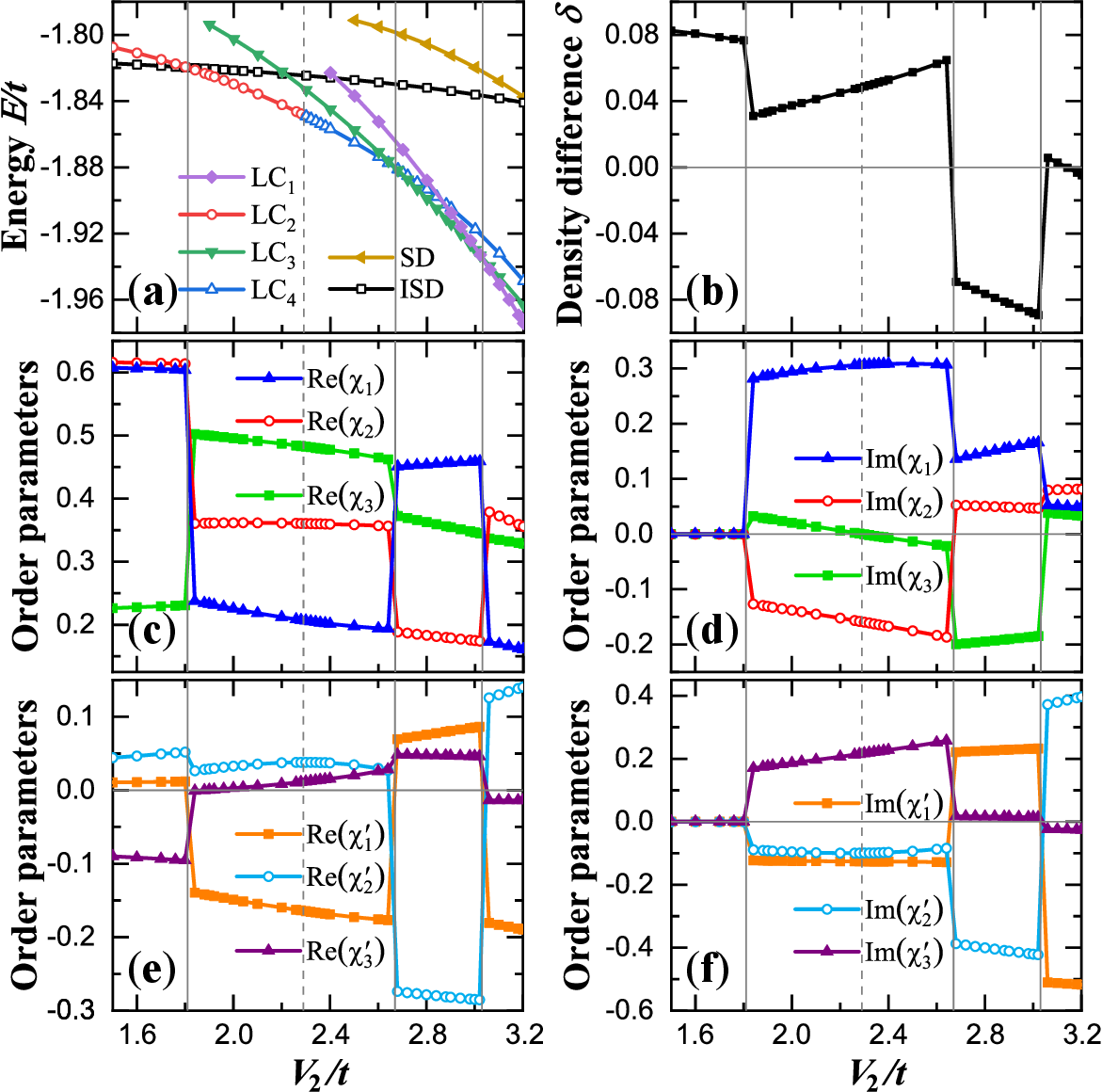}
\caption{Evolution of states as a function of nnn Coulomb repulsion $V_2$ at vH filling with fixed $V_1=1.75$. (a) Energy per site of all converged states. (b) The charge disproportionation $\delta$, (c,d) the real and imaginary part of nn bonds $\chiup_{1,2,3}$, and (e,f) the real and imaginary part of nnn bonds $\chiup'_{1,2,3}$ in the ground state with the lowest energy.}
\label{v2}
\end{center}
\end{figure}

To investigate in detail the phase transitions between different ground sates, we fix $V_1=1.75$ and monitor the phase evolution as a function of $V_2$ in the range from $1.5$ to $3.2$ that covers all of the five ground states displayed in Fig.~\ref{diagram}.
The results are reported in Fig.~\ref{v2}.
The energy evolution of the converged state is shown in Fig. \ref{v2}a, where if the endpoint of a curve is shown, the corresponding state is unstable beyond that point.
For instance, the orange line for the SD state ends at $V_2 \sim 2.5$, below which, calculations with any initial conditions could not converge self-consistently to a SD state.
The SD state obtained at $V_2 \gtrsim 2.5$ is much higher in energy and fails to present itself as a ground state in the phase diagram.
The charge disproportionation, i.e., the difference between electron densities on $A$ and $B$ sites $\delta=\frac{1}{2} (n_A - n_B)$, the nn bonds $\chiup_{1,2,3}$, and the nnn bonds $\chiup'_{1,2,3}$ in the ground states with the lowest energy are plotted in Figs \ref{v2}b-\ref{v2}f, respectively.

At small $V_2$, only ISD and LC$_2$ states can be obtained self-consistently, and the former is the ground state.
The inclusion of $V_2$ does not disturb the bond configuration in the real CDW state as $\chiup_1$ and $\chiup_2$ remain close in value in the ISD state.
As $V_2$ increases, the ground state changes from the topological trial ISD to the topological LC$_2$ Chern insulator via a first order transition at $V_2 \simeq 1.81$, as indicated by the level crossing in Fig.~\ref{v2}a and abrupt changes in the order parameters displayed in Fig. \ref{v2}b-\ref{v2}f.
We note that the insulating gap remains robust across the first order transition between the ISD and the LC states.
The real parts of the nn bonds in LC$_2$ plotted in Fig. \ref{v2}c now deviate strongly from the SD pattern due to the significant $V_2$, as $\text{Re}(\chiup_1)$ and $\text{Re}(\chiup_2)$ are much different in value.
The current on the  nn $\chiup_3$ bonds is much smaller than that on $\chiup_{1,2}$ bonds as shown in Fig.~\ref{v2}d.

Remarkably, increasing $V_2$ in the LC$_2$ phase, the small anticlockwise current on $\chiup_3$ decreases continuously and changes its sign at $V_2\simeq 2.29$ and turns into a clockwise current, as shown by the green curve in Fig.~\ref{v2}d.
The LC ground state thus changes from LC$_2$ illustrated in Fig.~\ref{states}b to LC$_4$ illustrated in Fig.~\ref{states}e, which are only different by the direction of the LC on the green nn bonds $\chiup_3$.
At the boundary between these two LC states, which is marked by the dashed thin line on the phase diagram Fig.~(\ref{diagram}), ${\rm Im}(\chiup_3)=0$ and the current on the green bonds vanishes.
Interestingly, the current pattern in this case consists of only closed loops, one encircling the center hexagon and the other around the two independent red triangles, forming an ideal vortex-antivortex pattern.
It is important to note that since LC$_2$ and LC$_4$ states has the same Chern number, and there is no discontinuity when crossing the boundary in the order parameters or the energy, these two LC states are in the same phase, just as between the other dotted thin lines in the phase diagram in Fig.~\ref{diagram} where the imaginary parts of the nnn bonds change signs.
Increasing $V_2$ further, successive first order phase transitions take place as the LC state enters the LC$_3$ and LC$_1$ phases accompanied by changes in the Chern number at $V_2 \simeq 2.67$ and $V_2 \simeq 3.03$, respectively.
We note in passing that the charge disproportionation (Fig. \ref{v2}b) is relatively small, $\delta/{\bar n} \lesssim 11\%$, over the parameter regimes in the phase diagram.

\begin{figure*}
\begin{center}
\fig{7in}{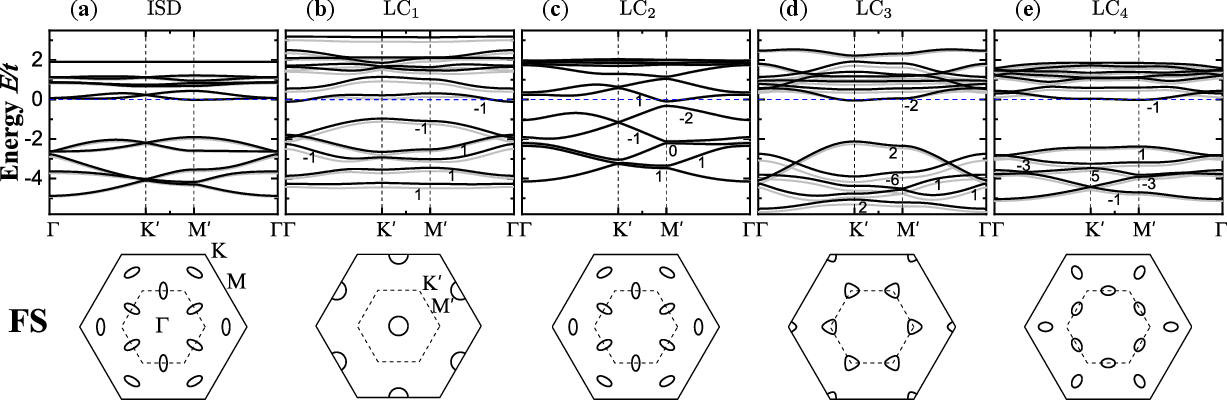}
\caption{Band dispersion (upper panels) and corresponding Fermi surfaces (lower panels) of the five ground states away from vH filling.
(a) ISD at $V=(2, 1)$,
(b) LC$_1$ at $V=(0.5, 2.5)$,
(c) LC$_2$ at $V=(0.8, 1.6)$,
(d) LC$_3$ at $V=(2, 3)$,
(e) LC$_4$ at $V=(2, 2.5)$.
Numbers denote the Chern numbers of corresponding bands.
The dispersions of the ground state at vH filling after a rigid shift are also plotted in grey lines for comparison.
Solid and dashed hexagons in lower panels denote, respectively, the original and the $2\times 2$ reduced BZ.
The self-consistently determined nn and nnn bond expectation values are listed in Table I.}
\label{awayvH}
\end{center}
\end{figure*}

\subsection{LC Chern metal away from vH filling}
We now turn to study the stability of the LC states and the nature of the ground state close to but away from vH filling. This is also directly relevant for the kagom\'e superconductors, since both LDA calculations \cite{Tan-prl21} and ARPES experiments \cite{Hu-xiv21, Nakayama-xiv21, Kang-xiv21, Luo-nc22} find that the actual Fermi level in \avs\ is slightly away from the vH singularity.
To this end, we set the band filling to $n =n_{\rm ed} =5.064/12$, which corresponds to the ``electron-doping" of the insulating CDW states, both the topological trivial and the topological orbital Chern insulators, discussed in the previous section.
For convenience of comparison, we investigate the self-consistent ground states at the same five sets of interaction parameters used in Fig. \ref{states} at the vH filling.
Remarkably, we find that each of the five ground states remains stable and turns into the corresponding metallic ground state upon small doping away from vH filling.
As shown in Fig. \ref{awayvH}, the obtained band structures only show small changes from those at vH filling, except for a rigid band shift.
The metallic CDW states near vH filling are thus characterized by the reconstructed FS consisting of small Fermi pockets around high-symmetry points in the reduced BZ.
These Fermi pockets are expected to be electron-like or hole-like depending on the location of the Fermi level relative to the vH singularity.
Interestingly, small-sized Fermi pockets due to the FS reconstruction induced by the CDW order have been detected recently by quantum oscillation experiments in the kagom\'e superconductors \cite{Wilson-prx21, Fu-prl21, Shrestha-prb22}.

We have shown that orbital Chern insulators at vH filling turn into LC Chern metals away from vH filling.
Fig.~\ref{awayvH} shows that the band crossing the Fermi level in each LC state carries a topological nontrivial Chern number.
The LC Chern metals are thus doped orbital Chern insulators with emergent CFPs.
Note that the LC states LC$_{1,3,4}$ only emerge under strong intersite Coulomb interactions with large energy gaps as shown in Fig. \ref{states} and Fig. \ref{awayvH}.
In contrast, the LC$_2$ state can be stabilized by moderately small interactions as can be seen in the phase diagram (Fig. \ref{diagram}) at vH filling.
This leads to a small energy gap that first opens at the M$^\prime$ points in the $2\times2$ folded zone, which we have verified numerically.
We thus argue that the LC$_2$ Chern metal in Fig.~\ref{awayvH}c with the CFPs pinned around the M$^\prime$ points may be the most relevant TRS breaking CDW state for the kagom\'e metals.

\begin{figure}
\begin{center}
\fig{3.4in}{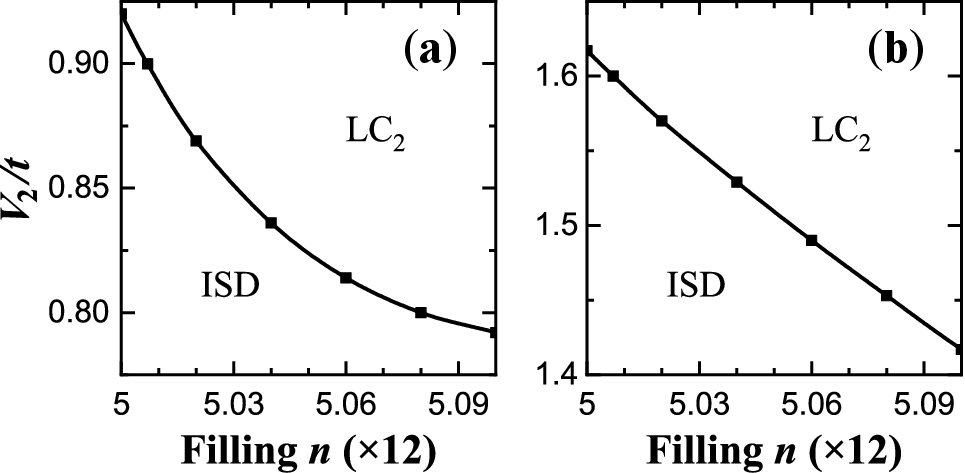}
\caption{Phase diagram in the plane spanned by band filling $n$ and nnn $V_2$, at fixed nn $V_1=0.5$ (a) and $V_1=1$ (b).}
\label{doping}
\end{center}
\end{figure}

To investigate further the effects of doping on LC$_2$, the state of interest, the phase diagram as a function of band filling $n$ and nnn $V_2$ is presented in Fig. \ref{doping} at fixed nn $V_1=$ 0.5 and 1.
Clearly, electron doping the kagom\'e lattice away from the vH filling helps the stabilization of LC$_2$ and reduces slightly the critical $V_2$ required for the emergence of loop currents.

\section{Summary and discussions}
We presented an effective model study of the TRS breaking CDW states on the kagom\'e lattice.
For the minimal single-orbital tight-binding model, we revealed that the susceptibility of the imaginary nnn bond exhibits a dominate divergences in the $2\times2$ breathing channel at vH filling.
Analyzing the leading instability of the vH modes, we found that the nnn Coulomb repulsion $V_2$ drives the instability toward the imaginary bond ordered CDW and hence the realization of LC, which competes with the instability toward the real bond CDW driven by the nn Coulomb interaction $V_1$.

This motivated us to carry out a fully self-consistent mean-field study of the concrete single-orbital $t$-$V_1$-$V_2$ model on kagom\'e lattice near vH fiilling.
We obtained the phase diagram at vH filling, and demonstrated that nn Coulomb repulsion $V_1$ alone drives the real ISD bond CDW state but unable to produce the LC states.
Remarkably, the nnn Coulomb interaction $V_2$ indeed induces the LC states as complex bond ordered CDWs neighboring the ISD state.
Four LC states with different current patterns appear in the phase diagram, which are identified as orbital Chern insulators by the topological properties of the quasiparticle band structures.
At the level of the self-consistent mean-field theory, the transitions between the topologically trivial ISD and the topological LC states are first order, without signatures of continuous gap-closing transitions and intermediate semimetal phases \cite{Lin-prb21}.

We found that electron doping away from the vH filling helps the stabilization of the $2\times2$ LC ordered states, as the doped orbital Chern insulators turn into LC Chern metals characterized by the CFPs due to the FS reconstruction.
In particular, since the LC$_2$ state arises from moderately small intersite Coulomb interactions in the phase diagram (Fig.~\ref{diagram}), we argue the doped LC$_2$ Chern metal with the CFPs centered around the M$^\prime$ points in the $2\times2$ folded zone (Fig.~\ref{awayvH}c) to be the most relevant TRS breaking state for kagom\'e metals.
The CFPs have been shown to carry concentrated Berry curvature and orbital magnetic moment with interesting experimental implications, including the large intrinsic contribution to the anomalous Hall conductivity \cite{SZ21}.
The outer CFPs along the $\Gamma$-K direction carry the dominant spectral weight and are connected by the wave vector $\frac{3}{4} \textbf{G}$ \cite{SZ21}.
They may play a crucial role in producing the $\frac{4}{3} \times \frac{4}{3}$ pair density waves as observed in the superconducting and pseudogap phases of the kagom\'e superconductor \cite{Chen-nat21}.

To end this paper, we comment on the effects of onsite Coulomb repulsion $U$ when included in the effective model.
If restricted to nonmagnetic states, the Hartree terms of onsite $U$ would only introduce density-dependent onsite potentials on each lattice sites, and suppress charge fluctuations.
We note that the charge disproportionation $\delta$ is relatively small over the entire parameter regime we explored.
As a result, onsite $U$ of reasonable strength is not expected to change qualitatively the phase diagram presented in Fig. \ref{diagram}.
It has been verified explicitly at $V=(2, 0)$ where ISD (ground state) has a larger charge disproportionation than LC$_2$ (metastalbe state).
A nonzero $U$ indeed reduce the energy difference between LC$_2$ and ISD.
However, at $U$ as large as $15$ where the charge disproportionation $\delta$ already become close in value in these two states, ISD is still substantially lower in energy and remains as the ground state.
This is consistent with previous results obtained for the $t$-$U$-$V_1$ model \cite{Kiesel-prl13, Wang-prb13} where a LC ground state is absent.
When magnetism is allowed, onsite $U$ would introduce long range magnetic orders at sufficient strength \cite{Kiesel-prl13, Wang-prb13}.
Therefore, a sizeable $U$ is expected to introduce magnetic ordered states in the phase diagram presented in Fig. \ref{diagram}.
It would be interesting to investigate the competition between magnetism and bond ordered CDW.

The correlated electronic structure and quantum states usually require more sophisticated analytical and numerical methods for more accurate description and improved understanding.
The findings presented here can be viewed as a starting point for further studies.

\section{Acknowledgments}
We thank Yun-Peng Huang, Hong-Chen Jiang, and are indebted to Chandra Varma for numerous invaluable discussions.
JD and SZ are supported by the National Key R\&D Program of China (Grant No. 2022YFA1403800), the Strategic Priority Research Program of CAS (Grant No. XDB28000000) and the National Natural Science Foundation of China (Grants No. 11974362 and No. 12047503).
ZW is supported by the U.S. Department of Energy, Basic Energy Sciences (Grant No. DE-FG02-99ER45747) and by the Research Corporation for Science Advancement (Cottrell SEED Award No. 27856).
ZW thanks Aspen Center for Physics for hospitality and acknowledges the support of NSF Grant No. PHY-1067611. Numerical calculations in this work were performed on the HPC Cluster of ITP-CAS.


\begin{thebibliography}{99}
\bibitem{Ortiz-prm19}
B. R. Ortiz, L. C. Gomes, J. R. Morey, M. Winiarski, M. Bordelon, J. S. Mangum, I. W. H. Oswald, J. A. Rodriguez-Rivera, J. R. Neilson, S. D. Wilson, E. Ertekin, T. M. McQueen, and E. S. Toberer, New kagome prototype materials: discovery of KV$_3$Sb$_5$, RbV$_3$Sb$_5$, and CsV$_3$Sb$_5$, Phys. Rev. Mater. {\bf3}, 094407 (2019).

\bibitem{Hasan-nm21}
Y.-X. Jiang, J.-X. Yin, M. M. Denner, N. Shumiya, B. R. Ortiz, G. Xu, Z. Guguchia, J. He, Md S. Hossain, X. Liu, J. Ruff, L. Kautzsch, S. S. Zhang, G. Chang, I. Belopolski, Q. Zhang, T. A. Cochran, D. Multer, M. Litskevich, Z.-J. Cheng, X. P. Yang, Z. Wang, R. Thomale, T. Neupert, S. D. Wilson, and M. Z. Hasan, Unconventional chiral charge order in kagome superconductor KV$_3$Sb$_5$, Nat. Mater. {\bf 20}, 1353 (2021).
\bibitem{Zeljkovic-nat21}
H. Zhao, H. Li, B. R. Ortiz, S. M. L. Teicher, T. Park, M. Ye, Z. Wang, L. Balents, S. D. Wilson, and I. Zeljkovic, Cascade of correlated electron states in the kagome superconductor CsV$_3$Sb$_5$, Nature {599}, 216 (2021).
\bibitem{Wilson-prx21}
B. R. Ortiz, S. M. L. Teicher, L. Kautzsch, P. M. Sarte, N. Ratcliff, J. Harter, J. P. C. Ruff, R. Seshadri, and S. D. Wilson, Fermi Surface Mapping and the Nature of Charge-Density-Wave Order in the Kagome Superconductor CsV$_3$Sb$_5$, Phys. Rev. X {\bf 11}, 041030 (2021).
\bibitem{XHChen-prx21}
Z. Liang, X. Hou, F. Zhang, W. Ma, P. Wu, Z. Zhang, F. Yu, J.-J. Ying, K. Jiang, L. Shan, Z. Wang, and X.-H. Chen, Three-Dimensional Charge Density Wave and Surface-Dependent Vortex-Core States in a Kagome Superconductor CsV$_3$Sb$_5$, Phys. Rev. X {\bf 11}, 031026 (2021).
\bibitem{Uykur-prb21}
E. Uykur, B. R. Ortiz, O. Iakutkina, M. Wenzel, S. D. Wilson, M. Dressel, and A. A. Tsirlin, Low-energy optical properties of the nonmagnetic kagome metal CsV$_3$Sb$_5$, Phys. Rev. B {\bf 104}, 045130 (2021).
\bibitem{Hasan-prb21}
N. Shumiya, Md. S. Hossain, J.-X. Yin, Y.-X. Jiang, B. R. Ortiz, H. Liu, Y. Shi, Q. Yin, H. Lei, S. S. Zhang, G. Chang, Q. Zhang, T. A. Cochran, D. Multer, M. Litskevich, Z.-J. Cheng, X. P. Yang, Z. Guguchia, S. D. Wilson, and M. Z. Hasan, Intrinsic nature of chiral charge order in the kagome superconductor RbV$_3$Sb$_5$, Phys. Rev. B {\bf 104}, 035131 (2021).
\bibitem{Zeljkovic-np22}
H. Li, H. Zhao, B. R. Ortiz, T. Park, M. Ye, L. Balents, Z. Wang, S. D. Wilson, and I. Zeljkovic, Rotation symmetry breaking in the normal state of a kagome superconductor KV$_3$Sb$_5$, Nat. Phys. {\bf 18}, 265 (2022).

\bibitem{Ortiz-prl20}
B. R. Ortiz, S. M. L. Teicher, Y. Hu, J. L. Zuo, P. M. Sarte, E. C. Schueller, A. M. Milinda Abeykoon, M. J. Krogstad, S. Rosenkranz, R. Osborn, R. Seshadri, L. Balents, J. He, and S. D. Wilson, CsV$_3$Sb$_5$: A Z$_2$ topological kagome metal with a superconducting ground state, Phys. Rev. Lett. {\bf 125}, 247002 (2020).
\bibitem{Ortiz-prm21}
B. R. Ortiz, P. M. Sarte, E. M. Kenney, M. J. Graf, S. M. L. Teicher, R. Seshadri, and S. D. Wilson, Superconductivity in the $\mathbb{Z}_2$ kagome metal KV$_3$Sb$_5$, Phys. Rev. Mater. {\bf 5}, 034801 (2021).
\bibitem{Yin-cpl21}
Q. Yin, Z. Tu, C. Gong, Y. Fu, S. Yan, and H. Lei, Superconductivity and Normal-State Properties of Kagome Metal RbV$_3$Sb$_5$ Single Crystals, Chin. Phys. Lett. {\bf 38}, 037403 (2021).
\bibitem{Chen-nat21}
H. Chen, H. Yang, B. Hu, Z. Zhao, J. Yuan, Y. Xing, G. Qian, Z. Huang, G. Li, Y. Ye, S. Ma, S. Ni, H. Zhang, Q. Yin, C. Gong, Z. Tu, H. Lei, H. Tan, S. Zhou, C. Shen, X. Dong, B. Yan, Z. Wang, and H.-J. Gao, Roton pair density wave in a strong-coupling kagome superconductor, Nature {\bf 599}, 222 (2021).

\bibitem{Kenney-jpcm21}
E. M. Kenney, B. R. Ortiz, C. Wang, S. D. Wilson, and M. J. Graf, Absence of local moments in the kagome metal KV$_3$Sb$_5$ as determined by muon spin spectroscopy, J. Phys.: Condens. Matter {\bf 33}, 235801 (2021).

\bibitem{Yang-sa20}
S.-Y. Yang, Y. Wang, B. R. Ortiz, D. Liu, J. Bayles, E. Derunova, R. Gonzalez-Hernandez, L. \v{S}mejkal, Y. Chen, S. S. P. Parkin, S. D. Wilson, E. S. Toberer, T. Mcqueen, Giant, unconventional anomalous Hall effect in the metallic frustrated magnet candidate, KV$_3$Sb$_5$, Sci. Adv. {\bf 6}, eabb6003 (2020).
\bibitem{Yu-prb21}
F. H. Yu, T. Wu, Z. Y. Wang, B. Lei, W. Z. Zhuo, J. J. Ying, and
X. H. Chen, Concurrence of anomalous Hall effect and charge
density wave in a superconducting topological kagome metal,
Phys. Rev. B {\bf 104} L041103 (2021).

\bibitem{nernest}
X. Zhou, H. Liu, W. Wu, K. Jiang, Y. Shi, Z. Li, Y. Sui, J. Hu, and J. Luo, Anomalous thermal Hall effect and anamalous Nernst effect of CsV$_3$Sb$_5$, Phys. Rev. B {\bf 105}, 205104 (2022).
\bibitem{seebeck}
D. Chen, B. He, M. Yao, Y. Pan, H. Lin, W. Schnelle, Y. Sun, J. Gooth, L. Taillefer, and C. Felser, Anomalous thermoelectric effects and quantum oscillations in the kagome metal CsV$_3$Sb$_5$, Phys. Rev. B {\bf 105}, L201109 (2022).

\bibitem{muSR1}
C. Mielke III, D. Das, J.-X. Yin, H. Liu, R. Gupta, Y.-X. Jiang, M. Medarde, X. Wu, H. C. Lei, J. Chang, P. Dai, Q. Si, H. Miao, R. Thomale, T. Neupert, Y. Shi, R. Khasanov, M. Z. Hasan, H. Luetkens, and Z. Guguchia, Time-reversal symmetry-breaking charge order in a kagome superconductor, Nature {\bf 602}, 245 (2022).

\bibitem{muSR2}
L. Yu, C. Wang, Y. Zhang, M. Sander, S. Ni, Z. Lu, S. Ma, Z. Wang, Z. Zhao, H. Chen, K. Jiang, Y. Zhang, H. Yang, F. Zhou, X. Dong, S. L. Johnson, M. J. Graf, J. Hu, H.-J. Gao, and Z. Zhao, Evidence of a hidden flux phase in the topological kagome metal CsV$_3$Sb$_5$, arXiv:2107.10714 (2021).

\bibitem{kerr}
Q. Wu, Z. X. Wang, Q. M. Liu, R. S. Li, S. X. Xu, Q. W. Yin, C. S. Gong, Z. J. Tu, H. C. Lei, T. Dong, and N. L. Wang, Simultaneous formation of two-fold rotation symmetry with charge order in the kagome superconductor CsV$_3$Sb$_5$ by optical polarization rotation measurement, Phys. Rev. B {\bf 106}, 205109 (2022).

\bibitem{kerr-liangwu}
Y. Xu, Z. Ni, Y. Liu, B. R. Ortiz, Q. Deng, S. D. Wilson, B. Yan, L. Balents, and L. Wu, Three-state nematicity and magneto-optical Kerr effect in the charge density waves in kagome superconductors, Nat. Phys. {\bf 18}, 1470 (2022).

\bibitem{kerr-yonezawa}
Y. Hu, S. Yamane, G. Mattoni, K. Yada, K. Obata, Y. Li, Y. Yao, Z. Wang, J. Wang, C. Farhang, J. Xia, Y. Maeno, and S. Yonezawa, Time-reversal symmetry breaking in charge density wave of CsV$_3$Sb$_5$ detected by polar Kerr effect, arXiv:2208.08036 (2022).

\bibitem{transport}
C. Guo, C. Putzke, S. Konyzheva, X. Huang, M. Gutierrez-Amigo, I. Errea, D. Chen, M. G. Vergniory, C. Felser, M. H. Fischer, T. Neupert, and P. J. W. Moll, Switchable chiral transport in charge-ordered kagome metal CsV$_3$Sb$_5$, Nature {\bf 611}, 461 (2022).

\bibitem{Varma}
C. M. Varma, Non-Fermi-liquid states and pairing instability of a general model of copper oxide metals, Phys. Rev. B {\bf 55}, 14554 (1997).

\bibitem{Varma-prl}
C. M. Varma, Pseudogap Phase and the Quantum-Critical Point in Copper-Oxide Metals, Phys. Rev. Lett. {\bf 83}, 3538 (1999).

\bibitem{staggeredflux}
I. Affleck and J. B. Marston, Large-$n$ limit of the Heisenberg-Hubbard model: Implications for high-$T_c$ superconductors, Phys. Rev. B {\bf 37}, 3774 (1988).

\bibitem{PALee-rmp2006}
P. A. Lee, N. Nagaosa, and X.-G. Wen, Doping a Mott insulator: Physics of high-temperature superconductivity, Rev. Mod. Phys. {\bf 78}, 17 (2006).

\bibitem{Nayak-prb2000}
C. Nayak, Density-wave states of nonzero angular momentum, Phys. Rev. B {\bf 62}, 4880 (2000).

\bibitem{ddw}
S. Chakravarty, R. B. Laughlin, D. K. Morr, and C. Nayak, Hidden order in the cuprates, Phys. Rev. B {\bf 63}, 094503 (2001).


\bibitem{JiangWang-arXiv22}
J. Ge, P. Wang, Y. Xing, Q. Yin, H. Lei, Z. Wang, and J. Wang,
Discovery of charge-4e and charge-6e superconductivity in kagome superconductor CsV$_3$Sb$_5$, arXiv:2201.10352 (2022).

\bibitem{Tan-prl21}
H. Tan, Y. Liu, Z. Wang, and B. Yan, Charge Density Waves and Electronic Properties of Superconducting Kagome Metals, Phys. Rev. Lett. {\bf 127}, 046401 (2021).
\bibitem{Wenzel-xiv21}
M. Wenzel, B. R. Ortiz, S. D. Wilson, M. Dressel, A. A. Tsirlin, and E. Uykur, Optical study of RbV$_3$Sb$_5$: Multiple density-wave gaps and phonon anomalies, Phys. Rev. B {\bf 105}, 245123 (2021).
\bibitem{Li-prx21}
H. Li, T. T. Zhang, T. Yilmaz, Y. Y. Pai, C. E. Marvinney, A. Said, Q. W. Yin, C. S. Gong, Z. J. Tu, E. Vescovo, C. S. Nelson, R. G. Moore, S. Murakami, H. C. Lei, H. N. Lee, B. J. Lawrie, and H. Miao, Observation of unconventional charge density wave without acoustic phonon anomaly in kagome superconductors AV$_3$Sb$_5$ (A = Rb, Cs). Phys. Rev. X {\bf 11}, 031050 (2021).

\bibitem{Ratcliff-prm21}
N. Ratcliff, L. Hallett, B. R. Ortiz, S. D. Wilson, and J. W. Harter, Coherent phonon spectroscopy and interlayer modulation of charge density wave order in the kagome metal CsV$_3$Sb$_5$, Phys. Rev. Mater. {\bf 5}, L111801 (2021).
\bibitem{Xie-prb22}
Y. Xie, Y. Li, P. Bourges, A. Ivanov, Z. Ye, J.-X. Yin, M. Z. Hasan, A. Luo, Y. Yao, Z. Wang, G. Xu, and Pengcheng Dai, Electron-phonon coupling in the charge density wave state of CsV$_3$Sb$_5$, Phys. Rev. B {\bf 105}, L140501 (2022).
\bibitem{Wu-prb22}
S. Wu, B. R. Ortiz, H. Tan, S. D. Wilson, B. Yan, T. Birol, and G. Blumberg, Charge density wave order in the kagome metal $A$V$_3$Sb$_5$ ($A=$ Cs, Rb, K), Phys. Rev. B {\bf 105}, 155106 (2022).
\bibitem{Liu-nc22}
G. Liu, X. Ma, K. He, Q. Li, H. Tan, Y. Liu, J. Xu, W. Tang, K. Watanabe, T. Taniguchi, L. Gao, Y. Dai, H.-H. Wen, B. Yan, and X. Xi, Observation of anomalous amplitude modes in the kagome metal CsV$_3$Sb$_5$, Nat. Commun. {\bf 13}, 3461 (2022).

\bibitem{Ferrari-prb22}
F. Ferrari, F. Becca, and R. Valent\'i, Charge-density waves in kagome-lattice extended Hubbard models at the van Hove filling, Phys. Rev. B {\bf 106}, L081107 (2022).

\bibitem{Park-prb21}
T. Park, M. Ye, and L. Balents, Electronic instabilities of kagome metals: Saddle points and Landau theory, Phys. Rev. B {\bf 104}, 035142 (2021).

\bibitem{Fernandes-xiv22}
M. H. Christensen, T. Birol, B. M. Andersen, and R. M. Fernandes, Loop Currents in $A$V$_3$Sb$_5$ kagome metals: multipolar and toroidal magnetic orders, Phys. Rev. B {\bf 106}, 144504 (2022).
\bibitem{Lin-prb21}
Y.-P. Lin and R. Nandkishore, Complex charge density waves at Van Hove singularity on hexagonal lattices: Haldane-model phase diagram and potential realization in kagome metals $A$V$_3$Sb$_5$, Phys. Rev. B {\bf 104}, 045122 (2021).
\bibitem{Feng-sb21}
X. Feng, K. Jiang, Z. Wang, and J. Hu, Chiral flux phase in the Kagome superconductor $A$V$_3$Sb$_5$, Science Bulletin {\bf 66}, 1384 (2021).
\bibitem{Feng-prb21}
X. Feng, Y. Zhang, K. Jiang, and J. Hu, Low-energy effective theory and symmetry classification of flux phases on the kagome lattice, Phys. Rev. B {\bf 104}, 165136 (2021).
\bibitem{Setty-xiv21}
C. Setty, H. Hu, L. Chen, and Q. Si, Electron correlations and T-breaking density wave order in a $Z_2$ kagome metal, arXiv:2105.15204 (2021).
\bibitem{Denner-prl21}
M. Denner, R. Thomale, and T. Neupert, Analysis of charge order in the kagome metal $A$V$_3$Sb$_5$ ($A$=K, Rb, Cs), Phys. Rev. Lett. {\bf 127}, 217601 (2021).
\bibitem{Yang-xiv22}
H.-J. Yang, H. S. Kim, M. Y. Jeong, Y. B. Kim, M. J. Han, and S. Lee, Intertwining orbital current order and superconductivity in Kagome metal, arXiv:2203.07365 (2022).
\bibitem{Mertz-npj22}
T. Mertz, P. Wunderlich, S. Bhattacharyya, F. Ferrari, and R. Valenti, Statistical learning of engineered topological phases in the kagome superlattice of AV$_3$Sb$_5$, npj Computational Materials {\bf 8}, 66 (2022).

\bibitem{SZ21}
S. Zhou and Z. Wang, Chern Fermi pocket, topological pair density
wave, and charge-4e and charge-6e superconductivity
in kagom\'e superconductors, Nat. Commun. {\bf 13}, 7288 (2022).

\bibitem{Kontani-xiv22}
R. Tazai, Y. Yamakawa, S. Onari, and H. Kontani, Mechanism of exotic density-wave and beyond-Migdal unconventional superconductivity in kagome metal AV$_3$Sb$_5$ (A = K, Rb, Cs), Sci. Adv. {\bf 8}, eabl4108 (2022).

\bibitem{Haldane}
F. D. M. Haldane, Model for a quantum Hall effect without Landau levels: condensed-matter realization of the parity anomaly, Phys. Rev. Lett. {\bf 61}, 2015 (1988).

\bibitem{Lin-prb22}
Y.-P. Lin and R. M. Nandkishore, Multidome superconductivity in charge density wave kagome metals. Phys. Rev. B {\bf 106}, L060507 (2022).

\bibitem{Kiesel-prl13}
M. L. Kiesel, C. Platt, and R. Thomale, Unconventional Fermi Surface Instabilities in the Kagome Hubbard Model, Phys. Rev. Lett. {\bf 110}, 126405 (2013).
\bibitem{Wu-prl21}
X.-X. Wu, T. Schwemmer, T. M\"{u}ller, A. Consiglio, G. Sangiovanni, D. Di Sante, Y. Iqbal, W. Hanke, A. P. Schnyder, M. M. Denner, M. H. Fischer, T. Neupert, and R. Thomale, Nature of unconventional pairing in the kagome superconductors $A$V$_3$Sb$_5$, Phys. Rev. Lett. {\bf 127}, 177001 (2021).

\bibitem{Wang-xiv21}
Z. Wang, S. Ma, Y. Zhang, H. Yang, Z. Zhao, Y. Ou, Y. Zhu, S. Ni, Z. Lu, H. Chen, K. Jiang, L. Yu, Y. Zhang, X. Dong, J. Hu, H.-J. Gao, and Z. Zhao, Distinctive momentum dependent charge-density-wave gap observed in CsV$_3$Sb$_5$ superconductor with topological Kagome lattice, arXiv:2104.05556 (2021).

\bibitem{Nakayama-xiv21}
K. Nakayama, Y. Li, M. Liu, Z. Wang, T. Takahashi, Y. Yao, and T. Sato, Multiple energy scales and anisotropic energy gap in the charge-density-wave phase of kagome superconductor CsV$_3$Sb$_5$, Phys. Rev. B {\bf 104}, L161112 (2021).

\bibitem{Hu-xiv21}
Y. Hu, S. M. L. Teicher, B. R. Ortiz, Y. Luo, S. Peng, L. Huai, J. Ma, N. C. Plumb, S. D. Wilson, J. He, and M. Shi, Charge-order-assisted topological surface states and flat bands in the kagome superconductor CsV$_3$Sb$_5$, Sci. Bull. {\bf 67}, 495 (2022).
\bibitem{Kang-xiv21}
M. Kang, S. Fang, J.-K. Kim, B. R. Ortiz, S. H. Ryu, J. Kim, J. Yoo, G. Sangiovanni, D. Di Sante, B.-G. Park, C. Jozwiak, A. Bostwick, E. Rotenberg, E. Kaxiras, S. D. Wilson, J.-H. Park, and R. Comin, Twofold van Hove singularity and origin of charge order in topological kagome superconductor CsV$_3$Sb$_5$, Nat. Phys. {\bf 18}, 301 (2022).

\bibitem{Luo-nc22}
H. Luo, Q. Gao, H. Liu, Y. Gu, D. Wu, C. Yi, J. Jia, S. Wu, X. Luo, Y. Xu, L. Zhao, Q. Wang, H. Mao, G. Liu, Z. Zhu, Y. Shi, K. Jiang, J. Hu, Z. Xu, and X. J. Zhou, Electronic nature of charge density wave and electron-phonon coupling in kagome superconductor KV$_3$Sb$_5$, Nat. Commun. {\bf 13}, 273 (2022).

\bibitem{Zhao-prb21}
J. Zhao, W. Wu, Y. Wang, and S. A. Yang, Electronic correlations in the normal state of kagome superconductor KV$_3$Sb$_5$, Phys. Rev. B {\bf 103}, L241117 (2021).

\bibitem{Wang-prb13}
W.-S. Wang, Z.-Z. Li, Y.-Y. Xiang, and Q.-H. Wang, Competing electronic orders on kagome lattices at van Hove filling, Phys. Rev. B {\bf 87}, 115135 (2013).

\bibitem{Raghu-prl08}
S. Raghu, X.-L. Qi, C. Honerkamp, and S.-C. Zhang, Topological Mott Insulators, Phys. Rev. Lett. {\bf 100}, 156401 (2008).

\bibitem{Zhu-prb13}
L. Zhu, V. Aji, and C. M. Varma, Ordered loop current states in bilayer graphene, Phys. Rev. B {\bf 87}, 035427 (2013).

\bibitem{Zhu-prb18}
Y. Ren, T.-S. Zeng, W. Zhu, and D. N. Sheng, Quantum anomalous Hall phase stabilized via realistic interactions on a kagome lattice, Phys. Rev. B {\bf 98}, 205146 (2018).

\bibitem{Zhu-prl16}
W. Zhu, S.-S. Gong, T.-S. Zeng, L. Fu, and D. N. Sheng, Interaction-Driven Spontaneous Quantum Hall Effect on a Kagome Lattice, Phys. Rev. Lett. {\bf 117}, 096402 (2016).


\bibitem{Jiang-prb16}
K. Jiang, J. Hu, H. Ding, and Z. Wang, Interatomic Coulomb interaction and electron nematic bond order in FeSe, Phys. Rev. B {\bf 93}, 115138 (2016).

\bibitem{Campo-JPCM10}
V. L. Campo Jr. and M. Cococcioni, Extended DFT+$U$+$V$ method with on-site and intersite electronic interactions, J. Phys.: Condens. Matter {\bf 22}, 055602 (2010).
\bibitem{Belozerov-prb12}
A. S. Belozerov, M. A. Korotin, V. I. Anisimov, and A. I. Poteryaev, Monoclinic $M_1$ phase of VO$_2$: Mott-Hubbard versus band insulator, Phys. Rev. B {\bf 85}, 045109 (2012).

\bibitem{Zhang-prb20}
Y. Zhang, K. Jiang, Z. Wang, and F.-C. Zhang, Correlated insulating phases of twisted bilayer graphene at commensurate filling fractions: a Hartree-Fock study, Phys. Rev. B {\bf 102}, 035136 (2020).

\bibitem{Xie-prl20}
M. Xie and A. H. MacDonald, Nature of the Correlated Insulator States in Twisted Bilayer Graphene, Phys. Rev. Lett. {\bf 124}, 097601 (2020).
\bibitem{Ashvin-prx20}
N. Bultinck, E. Khalaf, S. Liu, S. Chatterjee, A. Vishwanath, and M. P. Zaletel, Ground State and Hidden Symmetry of Magic-Angle Graphene at Even Integer Filling, Phys. Rev. X {\bf 10}, 031034 (2020).
\bibitem{Ashvin-prr21}
S. Liu, E. Khalaf, J. Y. Lee, and A. Vishwanath, Nematic topological semimetal and insulator in magic-angle bilayer graphene at charge neutrality, Phys. Rev. Research {\bf 3}, 013033 (2021).

\bibitem{Jiang-prl18}
K. Jiang, S. Zhou, X. Dai, and Z. Wang, Antiferromagnetic Chern Insulators in Noncentrosymmetric Systems, Phys. Rev. Lett. {\bf 120}, 157205 (2018).

\bibitem{Fu-prl21}
Y. Fu, N. Zhao, Z. Chen, Q. Yin, Z. Tu, C. Gong, C. Xi, X. Zhu, Y. Sun, K. Liu, and H. Lei, Quantum transport evidence of topological band structures of kagome superconductor CsV$_3$Sb$_5$,
Phys. Rev. Lett. {\bf 127}, 207002 (2021).

\bibitem{Shrestha-prb22}
K. Shrestha, R. Chapai, B. K. Pokharel, D. Miertschin, T. Nguyen, X. Zhou, D. Y. Chung, M. G. Kanatzidis, J. F. Mitchell, U. Welp, D. Popovic, D. E. Graf, B. Lorenz, and W. K. Kwok, Nontrivial Fermi surface topology of the kagome superconductor CsV$_3$Sb$_5$ probed by de Haas-van Alphen oscillations, Phys. Rev. B {\bf 105} 024508 (2022)

\end{thebibliography}
\end{document}